# Pressure effects on the magnetic structure in $La_{0.5}Ca_{0.5-x}Sr_xMnO_3$ ($0.1 \leq x \leq 0.4$) manganites

**Indu Dhiman[1], Thierry Strässle[2], L. Keller[2], B. Padmanabhan[2] and A. Das[1,*]**

[1]Solid State Physics Division, Bhabha Atomic Research Centre, Mumbai - 400085, India

[2]Laboratory for Neutron Scattering, ETH Zurich and Paul Scherrer Institute, CH - 5232 Villigen PSI, Switzerland

## Abstract

The effect of high pressure (0 – 8 GPa) on the magnetic structure of polycrystalline samples of $La_{0.5}Ca_{0.5-x}Sr_xMnO_3$ ($0.1 \leq x \leq 0.4$) manganites at 5 K is investigated using neutron diffraction technique. Application of pressure is found to modify the previously reported magnetic structure, observed under ambient conditions, in these compounds [I. Dhiman et al., Phys. Rev. B **77,** 094440 (2008)]. In x = 0.1 composition, at 4.6(2) GPa and beyond, A-type antiferromagnetic structure is found to coexist with CE-type antiferromagnetic phase, observed at ambient pressure, with $T_N \sim 150$ K. For x = 0.3 sample, as a function of pressure the CE-type phase is fully suppressed at 2.3(1) GPa and A-type antiferromagnetic phase is favored. Further Sr doping at x = 0.4, the A-type antiferromagnetic phase is observed at ambient pressure and for $T < T_N$ (~ 250K). This phase is retained in the studied pressure range. However, the magnetic moment progressively reduces with increasing pressure, indicating the suppression of A-type antiferromagnetic phase. The present study brings out the fragile nature of the CE-type antiferromagnetic state in half doped manganites as a function of pressure and disorder $\sigma^2$. We observe that pressure required for destabilizing the CE-type antiferromagnetic state is reduced with increasing disorder $\sigma^2$. External pressure and changing A-site ionic radii have analogous effect on the magnetic structure.

*E mail: *adas@barc.gov.in*

# I. INTRODUCTION

Perovskite manganites $R_{1-x}A_xMnO_3$ (R = trivalent rare earth ion and A = divalent ion Ca, Sr and Ba) exhibit wide variety of phenomena viz., colossal magnetoresistance, charge, spin and orbital ordering and mesoscopic phase separation. In particular, studies on half-doped perovskite manganites $R_{0.5}A_{0.5}MnO_3$, show a complex interplay of ferromagnetic double exchange, antiferromagnetic superexchange interactions, coupled to lattice distortions effects and orbital degrees of freedom which leads to a complex magnetic phase diagrams. The magnetic states of these compounds are influenced by external pressure, magnetic field and disorder ($\sigma^2$) as a result of doping at the rare earth site and/or transition metal site.[1-4] The disorder is quantified by variance of the ionic radii and is given as, $\sigma^2 = \sum x_i r_i^2 - \langle r_A \rangle^2$, where $x_i$ denotes the fractional occupancy of A-site ion and $r_i$ is the corresponding ionic radius.[5,6]

The conflicting nature of the effect of external pressure and disorder on the order of magnetic phase transition in ferromagnetic manganites has been reported recently.[7] It is found that pressure required for changing the first- order transition to second decreases as $\sigma^2$ increases. The effectiveness of the electronic localization effect also decreases with increasing $\sigma^2$. Recent pressure dependent study on electron doped manganite $Sm_{0.1}Ca_{0.9-y}Ba_yMnO_3$ (y = 0.02 and 0.06) showed the strong suppression of the ferromagnetic phase for the y = 0.06 sample, in comparison to y = 0.02 sample. [8] This behavior is attributed to the effect of chemical disorder $\sigma^2$, playing a more significant role than the effect related to the changes in $\langle r_A \rangle$. Pressure induced metallization behavior has been observed in ferromagnetic insulating compound, $La_{0.79}Ca_{0.21}MnO_3$.[9] Similarly, in undoped $LaMnO_3$ compound, metallization is observed at P ~ 32 GPa and suppression of orbital ordering occurs at P ~ 18GPa, which is attributed to suppression of Jahn-Teller distortions.[10]

There exist very few reports on the influence of pressure on the nature of magnetic structure in half doped compounds, particularly using neutron diffraction techniques. Half doped compounds can be broadly classified under low electron (low $\langle r_A \rangle$ regime) and high electron (high $\langle r_A \rangle$ regime) bandwidth (W) categories.[11] $Pr_{0.5}Sr_{0.5}MnO_3$ ($\langle r_A \rangle$ = 1.245 Å) compound exhibits a layered A-type antiferromagnetic structure, which is stable at ambient pressure and low temperature ($T < T_N$). The region of stability of this phase is enhanced at high pressure with an increase in $T_N$ and a phase separated region of orthorhombic A-type antiferromagnetic and tetragonal phase without long range order is found.[12] The $Nd_{0.5}Sr_{0.5}MnO_3$ ($\langle r_A \rangle$ = 1.237 Å) compound exhibits charge and orbitally ordered CE-type (charge exchange) antiferromagnetic insulating ground state. From pressure dependent resistivity studies it is conjectured that the CE-type AFM state is suppressed with pressure ~ 3.5 GPa in favor of an A-type antiferromagnetic state.[13] Similarly, pressure dependent resistivity studies raised the possibility that on application of hydrostatic pressure in $(Nd_{1-z}La_z)_{0.5}Sr_{0.5}MnO_3$ (z = 0.4) compound, a transformation of the CE-type to A-type antiferromagnetic spin structure would occur.[14] In A-site disordered manganites $Ln_{0.5}Ba_{0.5}MnO_3$ (Ln = Sm and Nd) the spin glass insulating state observed at ambient pressure, switches to a ferromagnetic metal state with increasing pressure, followed by a rapid increase of Curie temperature.[15] X-ray diffraction study in systems such as $Nd_{0.5}Ca_{0.5}MnO_3$ (with low $\langle r_A \rangle$ = 1.1715 Å) show that increase in pressure leads to development of shear strain in these compounds. The increase in shear strain with increasing pressure provides a mechanism for the insulating behavior of manganites at high pressures.[16] The compounds like $Nd_{0.5}Sr_{0.5}MnO_3$ and $La_{0.5}Ca_{0.5}MnO_3$ lie at the phase boundary of competing CE-type antiferromagnetic and ferromagnetic states. Therefore, study of these compounds is expected to shed more light on the pressure induced phase separation behavior in these compounds. Pressure induced phase separation behavior has been reported recently on doped bilayer manganites using x-ray diffraction studies.[17] Previous study on application of pressure on $La_{0.5}Ca_{0.5}MnO_3$ compound at 300 K, exhibit a structural transition from orthorhombic to monoclinic phase at 15 GPa. However, the CE type antiferromagnetic phase was found to be stable up to a maximum applied pressure of 6.2 GPa.[18]

In this communication, we show that increase in $\sigma^2$, makes the CE-type antiferromagnetic structure in $La_{0.5}Ca_{0.5}MnO_3$ compound susceptible to pressure. Above a critical pressure the CE type structure coexists with A-type antiferromagnetic order giving evidence of pressure induced phase separation behavior in these compounds. The critical pressure decreases with increasing $\sigma^2$. Towards this we have carried out neutron diffraction experiments on $La_{0.5}Ca_{0.5-x}Sr_xMnO_3$ ($0.1 \leq x \leq 0.4$) compounds at 5 K and pressures between 0 – 8 GPa. Substituting Ca with Sr leads to increase in the A-site ionic radii ($\langle r_A \rangle$) and disorder ($\sigma^2$). The magnetic phase diagram of these compounds at ambient pressure has been reported previously.[19] Under ambient condition the CE-type antiferromagnetic structure is stable for $x < 0.3$. At $x = 0.3$, CE-type ordering coexists with an A-type antiferromagnetic phase. The A-type antiferromagnetic order replaces the CE-type ordering at composition $x = 0.4$. All the compounds having composition in the range $0.1 \leq x \leq 0.4$, exhibit an insulating behavior. However, the resistivity of x = 0.4 at low temperatures is lower as compared to samples with x ≤ 0.3. Finally, at x = 0.5 composition the system becomes ferromagnetic and metallic. In the present high pressure study, we have chosen three compositions with different magnetic structures and have tried to understand the influence of external pressure on them. However, in the absence of high pressure synchrotron studies on these compounds we are unable to comment about the role of hydrostatic pressure on chemical structure in this study.

## II. **EXPERIMENT**

The polycrystalline samples $La_{0.5}Ca_{0.5-x}Sr_xMnO_3$ ($0.1 \leq x \leq 0.4$) used for this study are the same as the ones on which previous investigations were carried out.[19] Neutron diffraction data under high pressure were recorded on the DMC diffractometer (λ = 2.4575 Å), at the Swiss spallation neutron source (SINQ), Paul Scherrer Institute, Switzerland.[20] Pressure in the sample is generated by toroidal anvils made up of boron nitride.[21] Due to the strong absorption of natural boron, the anvils act as a beam stop for neutrons. The sample was loaded in the TiZr alloy gasket using a deuterated 4:1 methanol – ethanol mixture as a pressure transmitting medium. These gaskets accommodate a sample volume of 30-100 $mm^3$, depending on the maximum sought pressure. Since the TiZr alloy gaskets having average scattering length zero were used there were no additional Bragg reflections other than those from the sample and minute quantity of Pb used as pressure calibrant. The pressure was calibrated using the shift of Pb peaks, where Pb powder is mixed with the sample. The measurements were performed in the pressure range of 0 – 8.0 GPa and at temperature 5 K by placing the pressure cell in a helium cryostat. The time needed to obtain one spectrum was typically ~ 14 – 18 hrs. The diffraction data are analyzed by the Rietveld method using FULLPROF program.[22]

## III. RESULTS AND DISCUSSION

The Sr doped compounds in the series $La_{0.5}Ca_{0.5-x}Sr_xMnO_3$, for $x \leq 0.3$ are isostructural, having an orthorhombic distorted perovskite structure with *Pnma* space group. The x = 0.4 crystallizes with two orthorhombic phases in the space group *Pnma* and *Fmmm*. The structural and magnetic properties of these samples at ambient pressure have been reported previously.[19] The cell parameters obtained from Rietveld refinement of the diffraction data at high pressure are given in table I. The positional parameters of the atoms in the cell, bond lengths and bond angles could not be determined with sufficient accuracy due to the limited angular range available on the diffractometer and the inherent broadening of the Bragg reflections with increase in pressure. Therefore, these are not included in the table. Pressure induced orthorhombic to monoclinic structural transformations have been observed[18] in $La_{0.5}Ca_{0.5}MnO_3$ compound at very high pressure (P ~ 15 GPa) using synchrotron techniques. Since this is much beyond the maximum pressure employed in this study, pressure induced structural transitions can be ruled out in these samples.

The pressure dependence of lattice parameters at 5 K for $La_{0.5}Ca_{0.4}Sr_{0.1}MnO_3$ is shown in figure 1. The lattice parameters and volume reduces with increasing pressure. The calculated linear compressibility $k_i = -(1/a_{i0})(da_i/dP)_T$ ($a_i$ = a, b, and c) values for x = 0.1 composition at T = 5 K, for lattice parameters a, b and c are $k_a = 0.0017$ GPa$^{-1}$, $k_b = 0.0017$ GPa$^{-1}$ and $k_c = 0.0015$ GPa$^{-1}$, respectively. No significant change in lattice compressibilities along the three axes is observed. For x = 0.3 sample, $k_a = 0.0003$ GPa$^{-1}$ is smaller than that of the linear compressibilities along b and c - axis, $k_b = 0.0023$ GPa$^{-1}$ and $k_c = 0.0025$ GPa$^{-1}$, respectively. Similarly, for x = 0.4 sample, the values of $k_a$, $k_b$ and $k_c$ are 0.0015 GPa$^{-1}$, 0.0028 GPa$^{-1}$ and 0.0018 GPa$^{-1}$, respectively. The linear compressibility along the b - axis in x = 0.3 and 0.4 compound at 5 K are found to be similar to those reported in compounds exhibiting A-type antiferromagnetic order such as$Pr_{0.75}Na_{0.25}MnO_3$, $La_{0.75}Ca_{0.25}MnO_3$ and $Pr_{0.8}Na_{0.2}MnO_3$ compounds[23-25], suggesting similar nature of orbital ordering. Figure 2 and 3 displays the pressure dependence of orthorhombic distortions $Os_{\parallel}$ and $Os_{\perp}$ for x = 0.1 and 0.3 samples, respectively. To describe the orthorhombic distortions, Meneghini et al. defined the $Os_{\parallel} = 2\left(\frac{c-a}{c+a}\right)$ distortion in the ac plane and $Os_{\perp} = 2\left(\frac{a+c-b\sqrt{2}}{a+c+b\sqrt{2}}\right)$ along the b axis.[24] For x = 0.1 sample, the values of $Os_{\parallel}$ and $Os_{\perp}$ remain nearly constant at 0.012 and 0.014, respectively for all the pressures studied (2.6, 4.6 and 6.0 GPa), as shown in figure 2. At x =

0.3 composition, the $Os_{\|}$ is reduced as compared to x = 0.1 sample. In figure 3 the $Os_{\|}$ and $Os_{\perp}$ as a function of pressure for x = 0.3 is displayed. The $Os_{\|}$ exhibits a decrease as a function of pressure (0.0084 at P = 2.3 GPa to 0.00018 at P = 6.2 GPa), corresponding to reduction of strains in ac-plane. However, no significant change in $Os_{\perp}$ (0.015 at P = 2.3 GPa to 0.016 at P = 6.2 GPa) as a function of pressure is observed. Similar behavior of orthorhombic distortion was also observed for $Nd_{0.5}Sr_{0.5}MnO_3$ compounds.[26] This behavior is in contrast with parent compound $La_{0.5}Ca_{0.5}MnO_3$, in which case increase in $Os_{\|}$ is observed to increase and exhibits a slope change at ~ 5 GPa, while the $Os_{\perp}$ increases slightly with increasing pressure.[18]

Figure 4 displays the neutron diffraction patterns for $La_{0.5}Ca_{0.4}Sr_{0.1}MnO_3$ (x = 0.1) compound at temperature 5 K and pressure values 0.6(2) GPa, 4.6(2) GPa and 8.0(4) GPa. At ambient pressure and low temperature ($T < T_N$) this sample has CE-type antiferromagnetic structure. This is evidenced by antiferromagnetic superlattice reflections indexed as (0 1 ½) and (½ 1 ½), characteristic of CE-type antiferromagnetic ordering, which are observed below the antiferromagnetic transition temperature ($T_N \approx 150$ K). These superlattice reflections are indexed on a 2a×b×2c cell, having a CE-type structure, as proposed by Wollan and Kohler for the parent compound.[27] The CE-type antiferromagnetic structure consists of two Mn sublattices formed by ordered $Mn^{3+}$ and $Mn^{4+}$ ion, with propagation vectors $q_1$ = (0 0 ½) and $q_2$ = (½ 0 ½), respectively.[28] In CE-type model the spins are arranged as a one-dimensional zigzag chain of ferromagnetically coupled Mn spins in the *ac* plane and the chains in turn are coupled antiferromagnetically to each other.[29] On application of pressure, P = 0.6(2) GPa, the CE-type antiferromagnetic state is retained. However, the magnetic moment values at $Mn^{3+}$ and $Mn^{4+}$ sites are reduced to ~ 1.15 $\mu_B$/Mn at P = 0.6(2) GPa in comparison to the moment values of ~ 1.5 $\mu_B$/Mn at ambient pressure. This indicates the reduction of CE-type antiferromagnetic phase. Further suppression of the CE-type antiferromagnetic phase is observed on increasing pressure to 4.6(2) GPa. In addition, a new superlattice reflection, marked with an asterisk (+) in figure 4, is also observed. Rietveld analysis of the diffraction data shows that this superlattice reflection corresponds to pressure induced onset of A-type antiferromagnetic state. This superlattice reflection is indexed on a × 2b × c cell. In the A-type antiferromagnetic structure Mn magnetic moments form ferromagnetic planes with an antiferromagnetic coupling between them. Its characteristic feature is the apical compression of the $MnO_6$ octahedra along the direction perpendicular to the ferromagnetic planes. The magnetic moment for this A-type antiferromagnetic phase is 0.43(7) $\mu_B$/Mn, oriented ferromagnetically in the ac-plane and antiferromagnetically coupled along the b-axis. The magnetic moment is much lower than the expected value of 3.5 $\mu_B$/Mn. On further increasing the pressure to 8.0(4) GPa, the moment value of the A-type ordered phase is increased to 0.87(5) $\mu_B$/Mn, indicating the favor of A-type antiferromagnetic phase in the sample. However, with increasing pressure no significant change in the moment value for CE-type phase is observed. The CE-type antiferromagnetic spin structure has a characteristic orbital ordering of the type $d(3x^2 - r^2)/d(3z^2 - r^2)$. Favoring of A-type antiferromagnetic spin state with pressure would also lead to the orbital ordering of $d(x^2 - z^2)$ type. However, from the present high pressure work the change in the nature of orbital ordering is not evident. The variation of magnetic structure with pressure in this compound is different from that reported for the parent compound, $La_{0.5}Ca_{0.5}MnO_3$.[18] In the parent compound application of pressure did not yield any significant change in the magnetic structure (for pressure $P \leq 6.2$ GPa). This suggests that increasing disorder leads to instability of the CE-type magnetic structure. In a previous study, we found that a very small replacement of La with Y leading to reduction in volume by 0.6% in $La_{0.5}Ca_{0.5}MnO_3$ totally suppresses the antiferromagnetic structure.[30]

In figure 5 we show the neutron diffraction patterns for x = 0.3 at 5 K and at various pressures. At ambient pressure, this sample shows a mixture of CE- and A-type antiferromagnetic phases. Above a critical pressure of 2.3(1) GPa, the CE-type phase is fully suppressed and A-type antiferromagnetic phase is stabilized. The magnetic moment for Mn ion in the A-type antiferromagnetic state at P = 2.3(1) GPa and T = 5K, is 2.04(2) $\mu_B$, lying in the ac plane. On further increasing the pressure to 6.2(3) GPa, no change in the A-type antiferromagnetic spin structure is observed, although the moment value reduces slightly. With higher disorder in this sample as compared to x = 0.1, increase in pressure is found to suppress the CE-type phase indicating an existence of critical pressure. The value of the critical pressure decreases with increase in disorder. Figure 6 shows the diffraction pattern for x = 0.4 sample at pressures of 5 and 6.5 GPa. At ambient pressure, A-type antiferromagnetic structure is observed below 250K (~ $T_N$). With increase in pressure up to 6.5(3) GPa, no change in the magnetic structure is observed. However, the refined antiferromagnetic moment is found to reduce with increase in pressure. The moment values are 2.85(6), 1.89(3), and 1.63(2) $\mu_B$/Mn ion at 0, 5.0(3), and 6.5(3) GPa, respectively. Reduction in the moment values with increasing pressure indicates the destabilization of A-type antiferromagnetic phase with pressure.

In $La_{0.5}Ca_{0.5-x}Sr_xMnO_3$ series, substituting $Ca^{2+}$ ($<r_{Ca}>$ = 1.18 Å) with the larger $Sr^{2+}$ ($<r_{Sr}>$ = 1.31 Å) ion, both $<r_A>$ and $\sigma^2$ increases. We observe that either increase in external pressure or substituting Ca with an ion having higher A-site ionic radius $<r_A>$ such as, Sr[19] or Ba[31] (internal pressure) leads to similar magnetic phase transformations. In both these cases, the initial CE-type antiferromagnetic structure transforms into a phase with coexisting CE and A-type antiferromagnetic structure. Further increase in Sr or Ba doping leads to suppression of CE-type ordering and enhances A-type ordering, finally leading to a ferromagnetic metallic phase. In ferromagnetic manganites, in general, increase in $<r_A>$ is found to enhance $T_C$. This has been identified with increase in one electron bandwidth, W as a result of increase in Mn-O-Mn bond angles.[32-35] However, in half doped manganites change in $<r_A>$ is also found to result in a complex magnetic phase diagram.[11] Therefore, change in W alone does not explain the change in the magnetic structure with pressure observed in these systems. Additionally, change in $<r_A>$ results in increase in disorder, $\sigma^2$. The effect of random substitution at the A-site results in disorder causing random fields at the atomic level at Mn sites.[36,37] This leads to fluctuations in the hopping parameter and exchange coupling, which destroys the charge ordered state and/or results in the coexistence of ferromagnetic and antiferromagnetic phases.[38] Monte Carlo based studies in half doped manganites show the existence of CE- and A-type antiferromagnetic, and ferromagnetic metallic phases in the proposed magnetic phase diagram.[36,37] In the limit of large Hund's coupling the stability of CE-type structure is identified with finite electron phonon coupling ($\lambda$, related to Jahn-Teller distortion and hopping transition) and antiferromagnetic coupling ($J_{AF}$) between the $t_{2g}$ spins. At a given temperature, varying the electron phonon coupling is found to result in transformation of the phase from CE-type antiferromagnetic to ferromagnetic metallic phase with an intermediate A-type antiferromagnetic phase. In another study, the effect of increase in pressure is found to lower the polaronic phase.[7] Therefore, varying external pressure or A-site ionic radii $<r_A>$ may be identified with tuning the electron phonon coupling shown in the theoretical results. Additionally, our work also shows coexistence of A-type and CE-type antiferromagnetic phases in a certain pressure regime indicating that sharp boundaries do not separate the individual phases unlike the results obtained from simulation studies.

## IV. CONCLUSION

The effect of high pressure (0 – 8 GPa) on the magnetic structure of $La_{0.5}Ca_{0.5-x}Sr_xMnO_3$ (0.1 ≤ x ≤ 0.4) manganites at 5 K, have been investigated using neutron diffraction technique. In x = 0.1 compound, the low temperature CE-type antiferromagnetic phase observed at ambient pressure and T < $T_N$ (~ 150K), is retained for P < 4.6(2) GPa. Beyond this pressure, superlattice reflections corresponding to A-type antiferromagnetic structure emerges with increasing pressure and coexists with CE-type phase. For x = 0.3 sample, the CE-type phase is fully suppressed and A-type antiferromagnetic phase is favored at and beyond 2.3(1) GPa. The stabilization A-type antiferromagnetic phase coincides with the reduction of distortions in ac-plane. Further Sr doping at x = 0.4, the A-type antiferromagnetic phase is observed at ambient pressure and T < $T_N$. This phase is retained in the studied pressure range. The magnetic moment progressively reduces with increasing pressure, indicating the suppression of A-type antiferromagnetic phase. With increasing Sr doping, the pressure required for destabilizing the CE-type antiferromagnetic state is reduced. Above a critical pressure, the CE type structure coexists with A-type antiferromagnetic order giving evidence of pressure induced phase separation behavior in these compounds. The critical pressure decreases with increasing $\sigma^2$. A similarity in behavior of the magnetic structures is observed on changing the ionic radii and external pressure.

## ACKNOWLEDGEMENT

Financial support provided to Indu Dhiman by the Department of Science and Technology (DST), India for carrying out the neutron scattering experiments at PSI, Switzerland is gratefully acknowledged.

## Figure Captions

**Figure 1:** Pressure dependence of unit cell parameters and volume for $La_{0.5}Ca_{0.4}Sr_{0.1}MnO_3$ (x = 0.1) sample at 5 K.

**Figure 2:** Orthorhombic strains $Os_{\|}$ in the ac plane and $Os_{\perp}$ along b axis as a function of pressure for $La_{0.5}Ca_{0.4}Sr_{0.1}MnO_3$ (x = 0.1) at 5 K.

**Figure 3:** Orthorhombic strains $Os_{\|}$ in the ac plane and $Os_{\perp}$ along b axis as a function of pressure for $La_{0.5}Ca_{0.2}Sr_{0.3}MnO_3$ (x = 0.3) at 5 K.

**Figure 4:** Neutron diffraction pattern of $La_{0.5}Ca_{0.4}Sr_{0.1}MnO_3$ (x = 0.1) sample at P = 0.6, 4.6, 8.0 GPa and T = 5 K. Continuous lines through data points are fitted lines to the chemical and magnetic structure described in the text. The symbol (*) indicates the CE-type antiferromagnetic reflection and (+) indicates reflection corresponding to the A-type antiferromagnetic ordering. The tick marks in the first, second, third and fourth rows correspond to positions of Bragg reflections in nuclear, Pb, A-type antiferromagnetic and CE-type antiferromagnetic phases, respectively. Data for P = 4.6 and 8.0 GPa have been shifted vertically for clarity.

**Figure 5:** Neutron diffraction pattern of $La_{0.5}Ca_{0.2}Sr_{0.3}MnO_3$ (x = 0.3) sample at P = 2.3, 4.3, 6.2 GPa and T = 5 K. Continuous lines through data points are the fitted lines to chemical and magnetic structure described in the text. The symbol (+) indicate reflections corresponding to the A-type antiferromagnetic phase. The tick marks in the first, second and third rows correspond to positions of Bragg reflections in nuclear, Pb and A-type antiferromagnetic phases, respectively. The curve at the bottom of both the plots is difference between observed and calculated intensities for P = 6.2 GPa. Data for P = 6.2 and 4.3 GPa have been shifted vertically for clarity.

**Figure 6:** Neutron diffraction pattern of $La_{0.5}Ca_{0.1}Sr_{0.4}MnO_3$ (x = 0.4) sample at P = 6.5 and 5.0 GPa and T = 5 K. Continuous lines through data points are fitted lines to the chemical and magnetic structure described in the text. The symbol (+) indicate reflections corresponding to the A-type antiferromagnetic phase. The tick marks in the first, second and third rows correspond to positions of Bragg reflections in nuclear, Pb and A-type antiferromagnetic phases, respectively. The curve at the bottom of both the plots is difference between observed

and calculated intensities for P = 6.5 GPa. Data for P = 6.5 GPa has been shifted vertically for clarity.

## Table Captions

**Table I.** Structural parameters obtained from Rietveld refinement of neutron diffraction pattern at various pressures and T = 5 K for $La_{0.5}Ca_{0.5-x}Sr_xMnO_3$ (x = 0.1, 0.3 and 0.4) samples. Asterisks (*) symbol indicates that the refined parameters for P = 0 GPa, at 17K have been taken from reference [19].

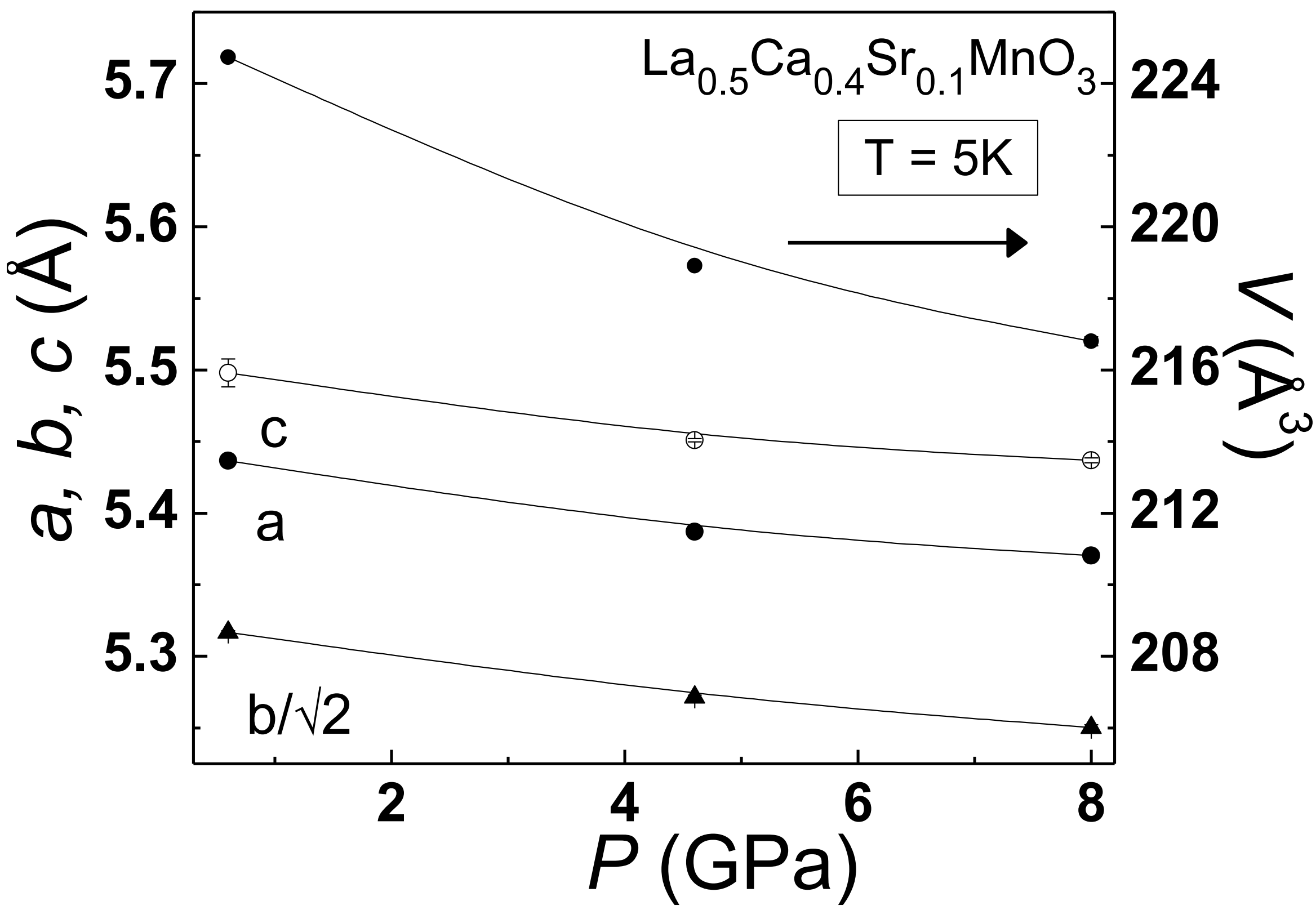

La0.5Ca0.4Sr0.1MnO3
T = 5K
a, b, c (Å)
V (Å3)
P (GPa)
c
a
b/√2
5.7
5.6
5.5
5.4
5.3
224
220
216
212
208
2
4
6
8

**Figure 1**

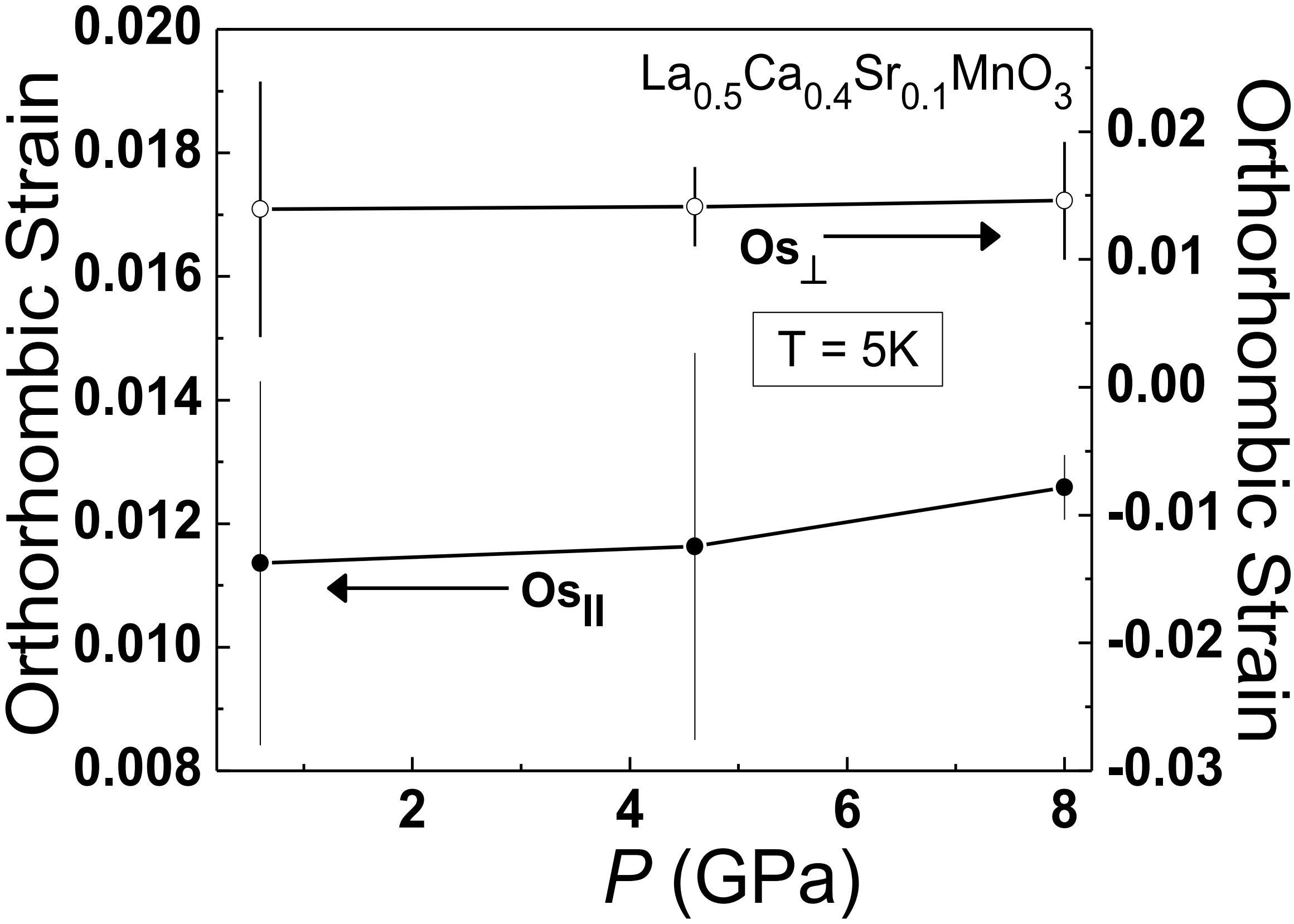

**Figure 2**

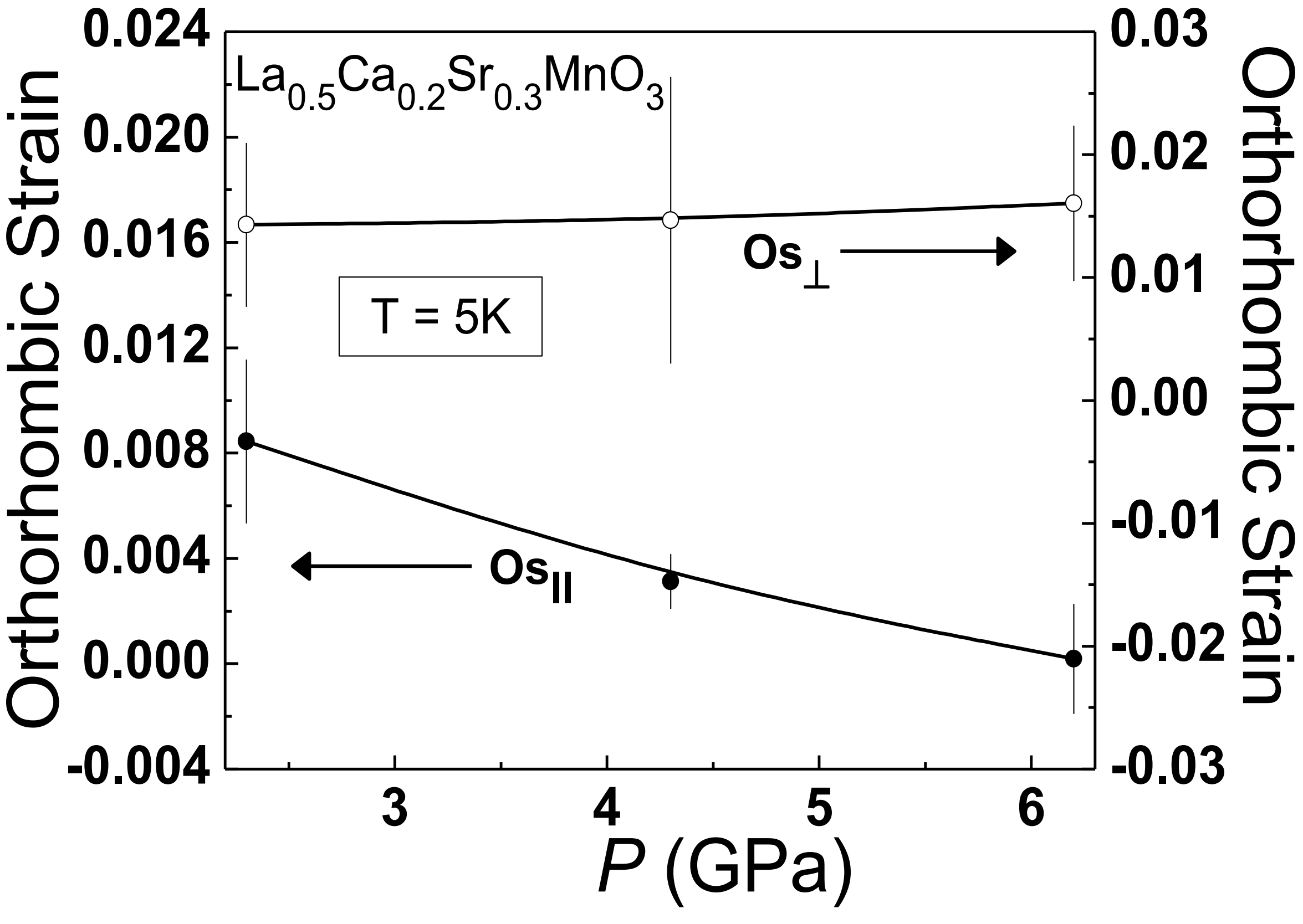

**Figure 3**

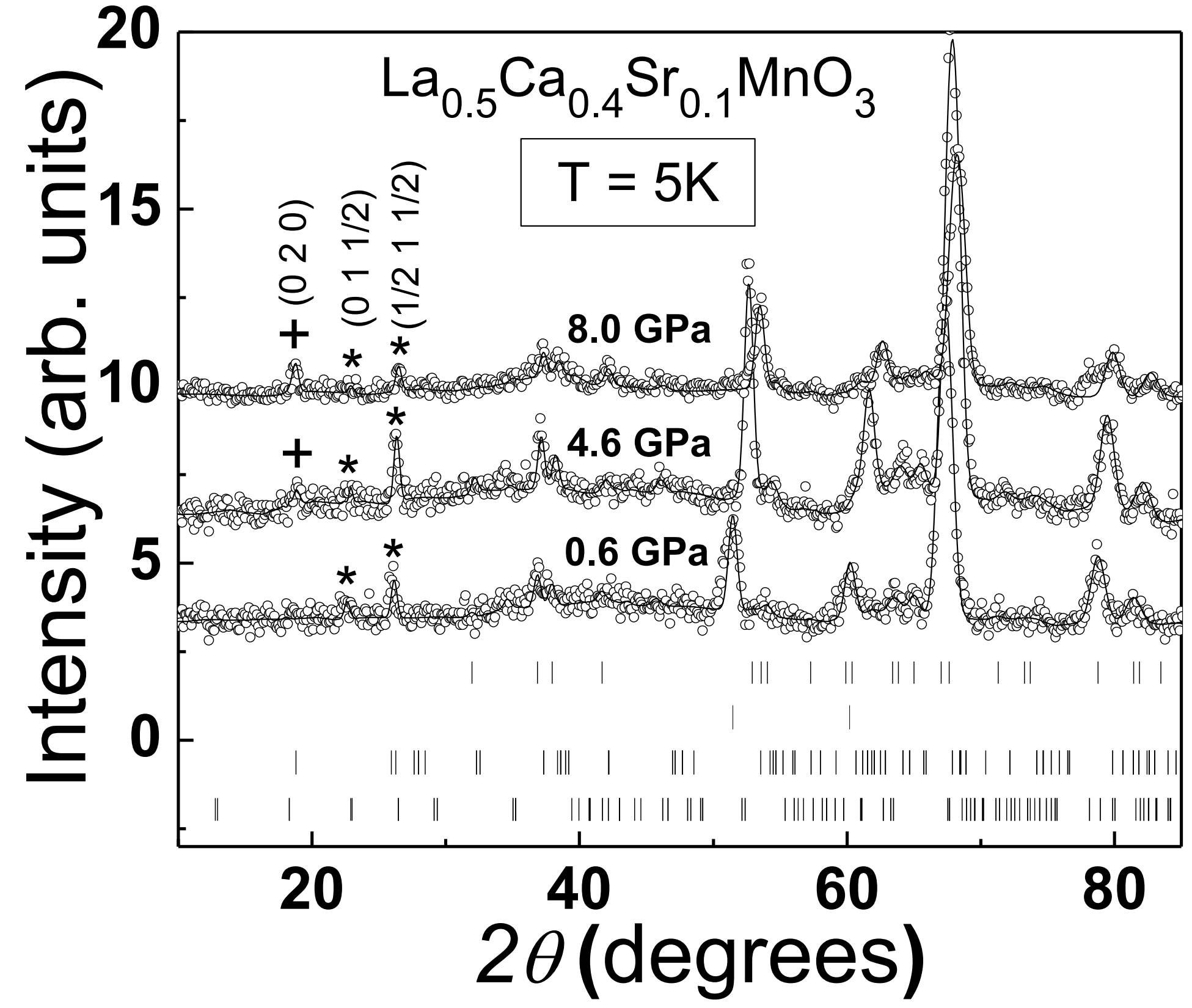

**Figure 4**

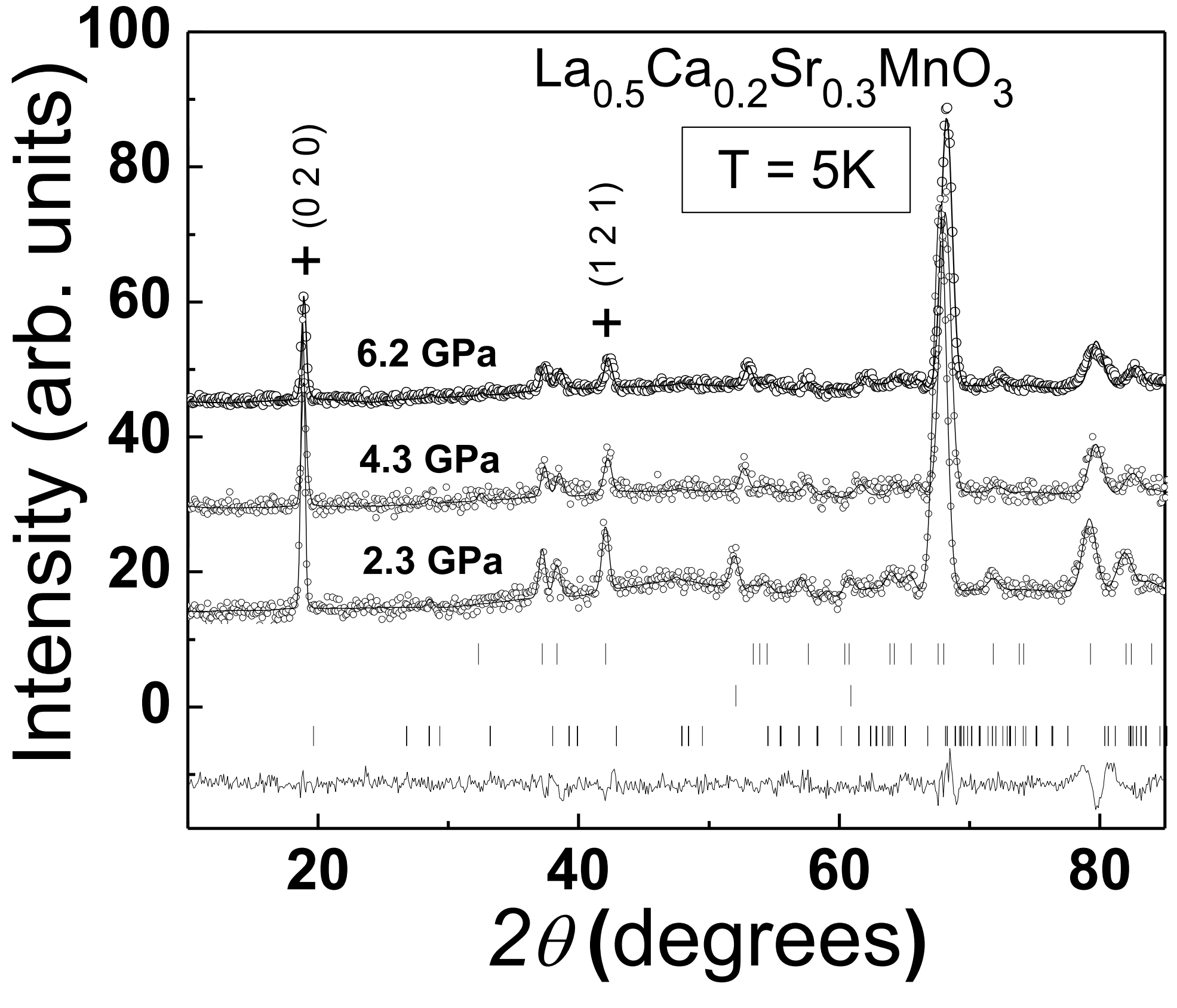

**Figure 5**

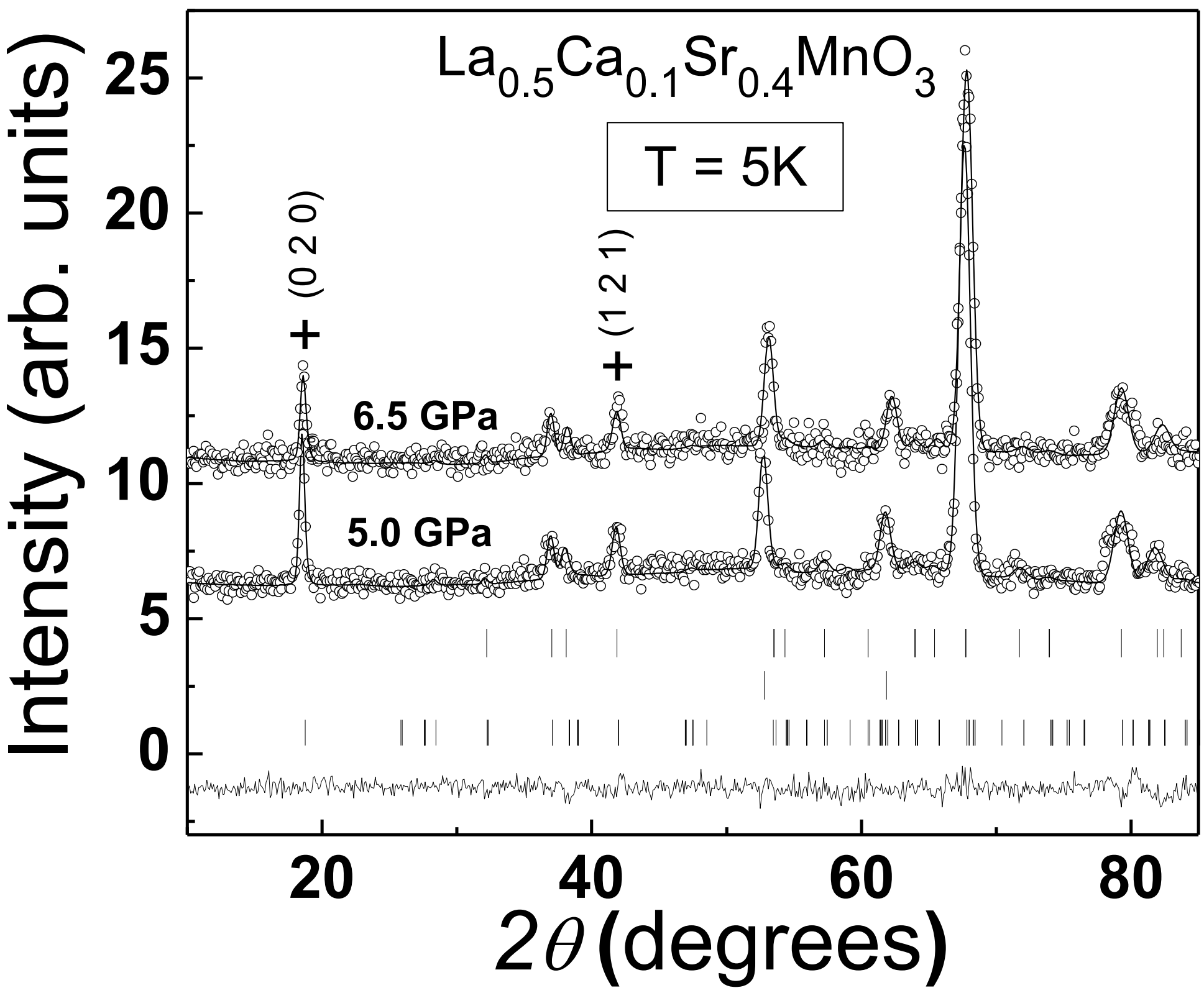

$La_{0.5}Ca_{0.1}Sr_{0.4}MnO_3$
T = 5K
(0 2 0)
(1 2 1)
6.5 GPa
5.0 GPa
Intensity (arb. units)
25
20
15
10
5
0
20
40
60
80
$2\theta$ (degrees)

**Figure 6**

Table I

| Refined parameters | x = 0.1 | | | | x = 0.3 | | | | x = 0.4 | | |
|---|---|---|---|---|---|---|---|---|---|---|---|
| **P (GPa)** | **0*** | **0.6 (2)** | **4.6 (2)** | **8.0 (4)** | **0*** | **2.3 (1)** | **4.3 (2)** | **6.2 (3)** | **0*** | **5.0 (3)** | **6.5 (3)** |
| **a (Å)** | 5.4426 (5) | 5.427 (1) | 5.386 (6) | 5.3601 (8) | 5.4406 (5) | 5.426 (6) | 5.422 (2) | 5.419 (4) | 5.447 (1) | 5.427 (6) | 5.415 (2) |
| **b (Å)** | 7.5347 (8) | 7.5069 (2) | 7.448 (2) | 7.409 (3) | 7.5289 (7) | 7.489 (2) | 7.458 (8) | 7.421 (2) | 7.547 (2) | 7.471 (5) | 7.4389 (8) |
| **c (Å)** | 5.4884 (6) | 5.489 (1) | 5.449 (6) | 5.428 (8) | 5.4937 (6) | 5.4720 (6) | 5.439 (2) | 5.418 (4) | 5.498 (1) | 5.431 (6) | 5.432 (2) |
| **V (Å$^3$)** | 225.07 (4) | 223.60 (8) | 218.70 (5) | 215.59 (6) | 225.03 (4) | 222.42 (4) | 220.5 (1) | 218.4 (2) | 226.00 (9) | 220.2 (3) | 218.8 (1) |
| **$M_{Mn^{3+}}$ CE-type AFM ($\mu_B$)** | 1.5 (1) | 1.15 (1) | 0.9 (1) | 1.0 (1) | 1.30 (4) | - | - | - | - | - | - |
| **$M_{Mn^{4+}}$ CE-type AFM ($\mu_B$)** | 1.5 (1) | 1.15 (7) | 1.26 (4) | 1.07 (7) | 1.30 (4) | - | - | - | - | - | - |
| **$M_{A\text{-type AFM}}$ ($\mu_B$)** | - | - | 0.43 (7) | 0.87 (5) | - | 2.04 (2) | 1.89 (3) | 1.63 (2) | 2.85 (6) | 1.89 (3) | 1.63 (2) |
| **$\chi^2$** | - | 1.26 | 1.41 | 1.76 | - | 1.24 | 1.12 | 1.94 | - | 1.69 | 1.47 |